\documentclass[10pt]{article}
\usepackage{graphicx}
\usepackage{amsfonts}
\begin{document}

\begin{titlepage}

\hfill{Preprint {\bf SB/F/01-289}}
\hrule
\vskip 2.5cm
\centerline{\bf On the Squeezed Number States and their Phase
Space Representations}
\vskip 2cm

\centerline{L. Albano $^{1}$, D.F.Mundarain $^{1}$ and J.
Stephany$^{1,2}$}
\vskip 4mm

\begin{description}

\item[] $^{1}${\it  Universidad Sim\'on
Bol\'{\i}var,Departamento de F\'{\i}sica,Apartado Postal 89000,  Caracas
1080-A, Venezuela.}
\item[] $^{2}${\it  Abdus Salam International Centre for Theoretical Physics,
Strada Costiera, 11, 34014  Trieste, Italy.}

\item[]{\it \ \ \ e-mail: lalbano@fis.usb.ve, dmundara@fis.usb.ve, stephany@usb.ve}
\end{description}
\vskip 1cm

{\bf Abstract}

\vskip 3mm

\noindent We compute the photon number distribution, the
$Q(\alpha)$ distribution function and the wave functions in the
momentum and position representation for a single mode squeezed
number state using  generating functions which allow to obtain any
matrix element in the squeezed number state representation from
the matrix elements in the squeezed coherent state representation.
For highly squeezed number states we discuss the previously
unnoted oscillations which appear in the $Q(\alpha)$ function. We
also note that these oscillations can be related to the
photon-number distribution oscillations and to the momentum
representation of the wave function.

\vskip 2cm \hrule
\bigskip
\centerline{\bf UNIVERSIDAD SIMON BOLIVAR} \vfill
\end{titlepage}

\section{Introduction}

\label{INTRO}

The study and experimental detection of squeezed states
\cite{Yuen1, Walls1, Wu1, Shelby1} and other non classical states
of optical systems, is an interesting issue, from both the
fundamental and the technological point of view. On one hand,
disentangling the properties of intrinsically quantum states of
light enhances our understanding of the behavior and interactions
of photons, and improves our knowledge of  Quantum
Electrodynamics. On the other hand, the use of light states with
reduced quantum fluctuations in one of the conjugate quadratures
is at the heart of attractive proposals for the detection of very
weak signals\cite{Caves1,Pace1}.

In this article we are concerned with the study of the squeezed
number states and their phase space representations . These states
emerged in the original analysis of Yuen \cite{Yuen1} of the Two
Photon Coherent States and have received since then a moderate
amount of attention in the literature \cite{Knight1,
Knight2,Gagen1,Nieto1,Dahl,Dariano1}. They appear promissory for
discussing the departure from the semiclassical regime in the
context of quantum optics and for this reason were chosen for this
study, but they also may have a practical interest on their own
\cite{Dariano1}.

The one mode squeezed number states are obtained by applying the
squeeze operator to the Fock states. As we discuss below, their
wave functions in the position and momentum like quadratures are
obtained by literally compressing or stretching the corresponding
wave functions of the Fock states. An important aspect which holds
our attention through this work is that they present oscillations
in the photon number distribution as have been reported in Refs.
\cite{Knight1,Knight2,Gagen1}. These oscillations were first
observed in the photon number distributions of squeezed states
\cite{Wheeler1} which are equivalent to the Two-Photon Coherent
States, and present several interesting properties. In a series of
papers, \cite{Wheeler1,Wheeler2,Walther1,Wheeler3, Schleich1},
these oscillations were explained as the result of the
interference of different contributions to probability amplitudes.
These contributions were interpreted as the overlapping of the
regions assigned to each quantum state on the phase space. For the
squeezed number states oscillations, a closely related discussion
was sketched in~\cite{Knight1,Knight2}. This treatment which is
physically very appealing is based on the semiclassical
representation of states as bounded areas in phase space obtained
with the help of Wigner distributions ~\cite{Wigner}. In Ref.
\cite{Wheeler3} for the case of the squeezed states this area was
fixed by weighting the Bohr-Sommerfeld rings with the Wigner-Cohen
\cite{Cohen1} distribution function. The same prescription was
followed in Refs. \cite{Knight1,Knight2,Gagen1} for the squeezed
number states. In Ref \cite{Dowling}, using the WKB method, a more
general framework for the use of this picture was established.

In this paper we study the properties of the squeezed number
states with the help of various representations. We compute
explicitly the probability amplitudes and distributions of the
squeezed number states in the Fock, position and momentum
representations. We also compute the $Q({\alpha})$ distribution
which has the advantage of being a true probability distribution
which can be interpreted as the probability of a simultaneous
measurement of position and momentum within some region of the
phase space. For high squeezing, the $Q(\alpha)$ function of a
squeezed number state  presents intrinsic oscillations in the
space of the complex labels $\alpha$. These oscillations were not
discussed previously in the literature and can be correlated with
those in the momentum and Fock representations.

Our computations are done using a generating function technique
which can be applied quite generally to compute expectation values
of arbitrary observables in different representations.

\section{Two photon coherent states and squeezed number states}
\label{TOCSINS}

We consider here a single mode of the radiation field described in
terms of the creation and annihilation operators $a^{\dagger}$,
$a$, the number operator $\hat{N}_{a}= a^{\dagger}a$, and the
momentum-like and position-like quadratures
$\hat{p}=\frac{1}{i\sqrt{2}}(a-a^{\dag})$ and
$\hat{q}=\frac{1}{\sqrt{2}}(a + a^{\dag})$. Number (Fock) states
$|n\rangle$ and coherent states $|\beta\rangle$ are the
eigenstates of $\hat{N}_{a}$ and $a$ with eigenvalues $n$ and
$\beta$ respectively. Coherent states may also be characterized as
displaced vacuum states and are minimal uncertainty states
satisfying $\Delta \hat{q}^{2}=\Delta \hat{p}^{2}=\frac{1}{2} $.
We are working in units such that $h=2\pi$. In the number state
representation they are written as:
\begin {equation}
\label{CSD1}
|\beta\rangle = e^{-|\beta|^2/2} \sum_n \frac{\beta^
n}{\sqrt{n!}}|n\rangle .
\end{equation}
Squeezed states and squeezed number states may be obtained
directly from $|\beta\rangle$ and $|n\rangle$ by the application
of the squeeze operator. This operator depends on a complex
parameter $\xi\equiv r e^{-i\phi}$. For simplicity we restrict our
computation to the case of real $\xi$. We define $S(r)$ as the
squeeze operator, and therefore
\begin{eqnarray}
|\beta,r\rangle = S(r) |\beta \rangle \label{eq:TRTCS}\\
|m,r\rangle = S(r) |m\rangle \label{eq:TRsqueezed number states}
\end{eqnarray}
where $S(r)$  is given by
\begin{equation} \label{eq:SQUEOP}
S(r) = \exp\left( \frac{ 1}{2}r(a^{\dagger})^2 - \frac{1}{2} ra^2\right).
\end{equation}
 It is also useful to define a transformed
annihilation operator $b$  as:
\begin{equation}\label{eq:BBOP}
b = S a S^{\dagger} = \cosh (r) a + \sinh (r) a^{\dagger}
\end{equation}
and then, one has
\begin{equation} \label{eq:EIGTCS}
b |\beta,r \rangle = \beta |\beta,r \rangle
\end{equation}
\noindent and
\begin{equation}
\hat{N}_{b} |m ,r \rangle = b^{\dag} b |m ,r \rangle = m |m ,r
\rangle .
\end{equation}

Given the linear nature of the transformation on $a$
(\ref{eq:BBOP}), one readily computes the coherent states and Fock
states amplitudes for the squeezed states~\cite{Yuen1},

\begin{eqnarray}
\langle\alpha|\beta,r\rangle = \frac{1}{\cosh^{1/2}(r)}
\exp \left(
  - \frac{1}{2}|\alpha|^2
  - \frac{1}{2}|\beta|^2
  -\frac{\tanh(r)}{2} \alpha^{* \, 2}\right.\nonumber\\
  \left.+\frac{\tanh(r)}{2} \beta^2
  +\frac{1}{\cosh(r)}\alpha^*\beta
\right) \label{eq:TCSCOH}
\end{eqnarray}
\begin{eqnarray}
\langle n |\beta,r\rangle = \frac{\tanh^{n/2}(r)}{(2^n
n! \cosh (r))^{1/2}} \mathrm{H}_{n}\left(\frac{\beta}{(2
\sinh (r) \cosh (r))^{1/2}}\right) \nonumber\\ \times \exp
\left( - \frac{1}{2}|\beta|^2
  +\frac{1}{2} \tanh(r)\beta^2 \right) \label{eq:TCSFOC}
\end{eqnarray}
\noindent where $\mathrm{H}_{n}$ is the $n$th order Hermite
polynomial.

For $\beta\;\;\epsilon\;\;\mathbb{R}$, the wave functions  in $q$
and $p$ are then given by:

\begin{equation} \label{eq:TCSPOS}
\langle q |\beta,\xi\rangle = (2 \pi \Delta^{2}
q)^{-1/4 }\exp \left(-\frac{(q-q_0)^2}{4\Delta^{2}
q}\right)
\end{equation}
\begin{equation} \label{eq:TCSMOM}
\langle p |\beta,\xi\rangle =(\frac{2 \Delta^{2}
q}{\pi})^{1/4 }\exp \left(-p^2 \Delta^{2}q -i p q_0
\right)\, ,
\end{equation}
with  $q_0 = \sqrt{2}\, e^r \beta
$, $\Delta^2 q = e^{-2 r} /2$ and
$\Delta^2p = e^{2 r} /2$.

\section{Representations of the squeezed number states}
\label{TCSSNS}

To discuss the properties of the squeezed number states, we
observe that for any operator $\hat{R}$ and any state
$|\psi\rangle$ , the completeness of the Fock states and Eq.
(\ref{CSD1}) allow one to write,
\begin{equation}
\label{eq:SERIETCS} \langle \psi|\hat{R}|\beta,r\rangle =\sum_{m}
\langle
 \psi|\hat{R}S(r)|m\rangle\langle m|\beta\rangle
= e^{- |\beta|^2/2} \sum_{m} \frac{\beta^m\langle
\psi|\hat{R}|m,r\rangle}{(m!)^{1/2}} .
\end{equation}

This provides a generating function for the matrix element
$\langle\psi|\hat{R}|m,r\rangle$ which can then be obtained as,

\begin{equation}
\label{eq:GENESNS}
\langle \psi|\hat{R}|m,r\rangle = \frac{1}{(m!)^{1/2}}\left[
\frac{\partial^m}{\partial\beta^m } ( e^{ |\beta|^2/2}\langle
\psi|\hat{R}|\beta,r\rangle) \right]_{\beta=0} .
\end{equation}

For example, if $\hat{R}=U(t)$ is the evolution operator, and
$|\psi\rangle=|k\rangle$ with $\{|k\rangle\}$ being any complete
basis of the radiation states, one can compute in this form the
time dependent probability amplitudes in the $\{|k\rangle\}$
representation. For $t=0$ we get the probability amplitudes of
stationary squeezed number states.

 Generalizing the same idea, we also note that
\begin{equation}
\langle \alpha_{g},r|\hat{R}|\beta_{g},r\rangle = e^{-1/2 | \alpha
| ^{2}} e^{-1/2 | \beta| ^{2}} \sum_{n}\sum_{m} \frac{\alpha^{* n}
}{(n!)^{1/2}} \frac{\beta^{m} }{(m!)^{1/2}}\langle n,r|\hat{R}|
m,r\rangle
\label{eq:DOPOISS}
\end{equation}
so that we can write,
\begin{equation}
\langle n,r|\hat{R}| m,r\rangle=\left.
\frac{1}{(n!m!)^{1/2}}\frac{\partial^{n}}{\partial
\alpha^{*n}}\frac{\partial^{m}}{\partial \beta^{m}}\left[
e^{1/2|\alpha|^{2}} e^{1/2|\beta|^{2}}\langle
\alpha_{g},r|\hat{R}|\beta_{g},r\rangle\right] \right |
_{\alpha^{*}=0,\beta=0} .
\label{eq:DOENTGEN}
\end{equation}

With this observation we  can generate the probability
distributions for the squeezed number states in the Fock,
position, momentum, and coherent states representations.   Taking
$|\psi\rangle\ = | n\rangle$ and $\hat R =1$, the amplitude for
the squeezed number states in the Fock states basis is,
\begin{eqnarray}
\langle n |m,r \rangle = \frac{1}{(m!)^{1/2}}\left[
\frac{\partial^m}{\partial\beta^m } ( e^{
|\beta|^2/2}\langle n |\beta_{g},r \rangle )
\right]_{\beta=0}\nonumber = \\
 \frac{(\tanh (r)/2)^{n/2}}{\sqrt{n! m!\cosh (r)}}
\left[ \frac{\partial^m}{\partial\beta^m }
\mathrm{H}_{n}\left(\frac{\beta}{\sqrt{2 \sinh (r) \cosh
(r)}} \right) e^{\frac{\beta^2\tanh
(r)}{2}}\right]_{\beta=0} .
\label{eq:GENFOCK}
\end{eqnarray}
Next we note that,
\begin{eqnarray}
\left[\mathrm{H}_n^{(l)}\left(\frac{\beta}{(2 \sinh (r)\cosh
(r))^{1/2}}\right)\right]_{\beta=0} \nonumber = \\
\frac{2^l (-1)^{n-l/2}n!(2\sinh (r) \cosh (r))^{-l/2}}{((n-l)/2)!}
\end{eqnarray}
and
\begin{eqnarray}
(e^{\frac{\beta^2\tanh(r)}{2}})^{(l)}|_{\beta=0} =\left\{\begin{array}{lc}
0 & l\mbox{ odd} \\
\left(\frac{\tanh(r)}{2}\right)^{l/2}\frac{l!}{(l/2)!}
& l\mbox{ even}
\end{array} \right. .
\end{eqnarray}
Using the Cauchy formula
\begin{equation}
\{ f(x) g(x)\}^{(m)} = \sum_{k=0}^m \frac{m!}{k!(m-k)!}
f^{(m-k)}(x) g^{(k)}(x) \label{eq:DERPRO}
\end{equation}
we get,
\begin{eqnarray}
\langle n | m,r \rangle =
\frac{(m!n!)^{1/2}}{\cosh(r)^{\frac{(n+m+1)}{2}}}
\sum^{\min(m,n)}_{k} (\frac{\sinh(r)}{2})^{\frac{(n+m-2k)}{2}}
\frac{(-1)^{\frac{(n-k)}{2}}}{k!(\frac{m-k}{2})!(\frac{n-k}{2})!}
\label{eq:SNSFOC}
\\ k=\left\{\begin{array}{lc} 0,2,4,6... & n,m\mbox{ even} \\
1,3,5,7... & n,m\mbox{ odd}
\end{array} \right. . \nonumber
\end{eqnarray}

This result is in agreement with the one reported in Ref.
\cite{Knight1}, which was obtained by the application of a
normally ordered squeeze operator to a Fock state. The most
salient features of this photon distribution are that it
oscillates and that only the photon number states of the same
parity of $m$ are represented in the expansion for $| m,r
\rangle$. The latter reflects of course the quadratic dependence
of $S(r)$ in $a$ and $a^\dagger$. The oscillations of the
distribution $P_{n,m}\equiv |\langle n | m,r \rangle|^{2}$ for
$m>1$ and $r$ sufficiently large have a fixed number of maxima;
$m/2+1$ when $m$ is even and $(m+1)/2$ when $m$ is odd. These
characteristics can be related to the structure of the $Q(\alpha)$
function, as we  discuss below.

For the squeezed number states with $m=0$ (the squeezed vacuum)
and $m=1$ the sum in equation (\ref{eq:SNSFOC}) is just a single
term:
 \begin{eqnarray}
   |\langle n| 0,r
\rangle|^{2}=\frac{n!}{((\frac{n}{2})!)^{2}2^{n}}\frac{\tanh^{n}(r)}{\cosh(r)}
\label{eq:ENCFOC0}\\
 n=0,2,4,6,.... \nonumber
 \end{eqnarray}
\begin{eqnarray}
|\langle n|1,r\rangle|^{2}=\frac{n!}{((\frac{n-1}{2})!)^{2}2^{n-1}}
\frac{\tanh^{n-1}(r)}{\cosh^{3}(r)} .
\label{eq:ENCFOC1}\\
 n=1,3,5,7,....
\nonumber
 \end{eqnarray}

\begin{figure}[pt]
\includegraphics[scale=0.5]{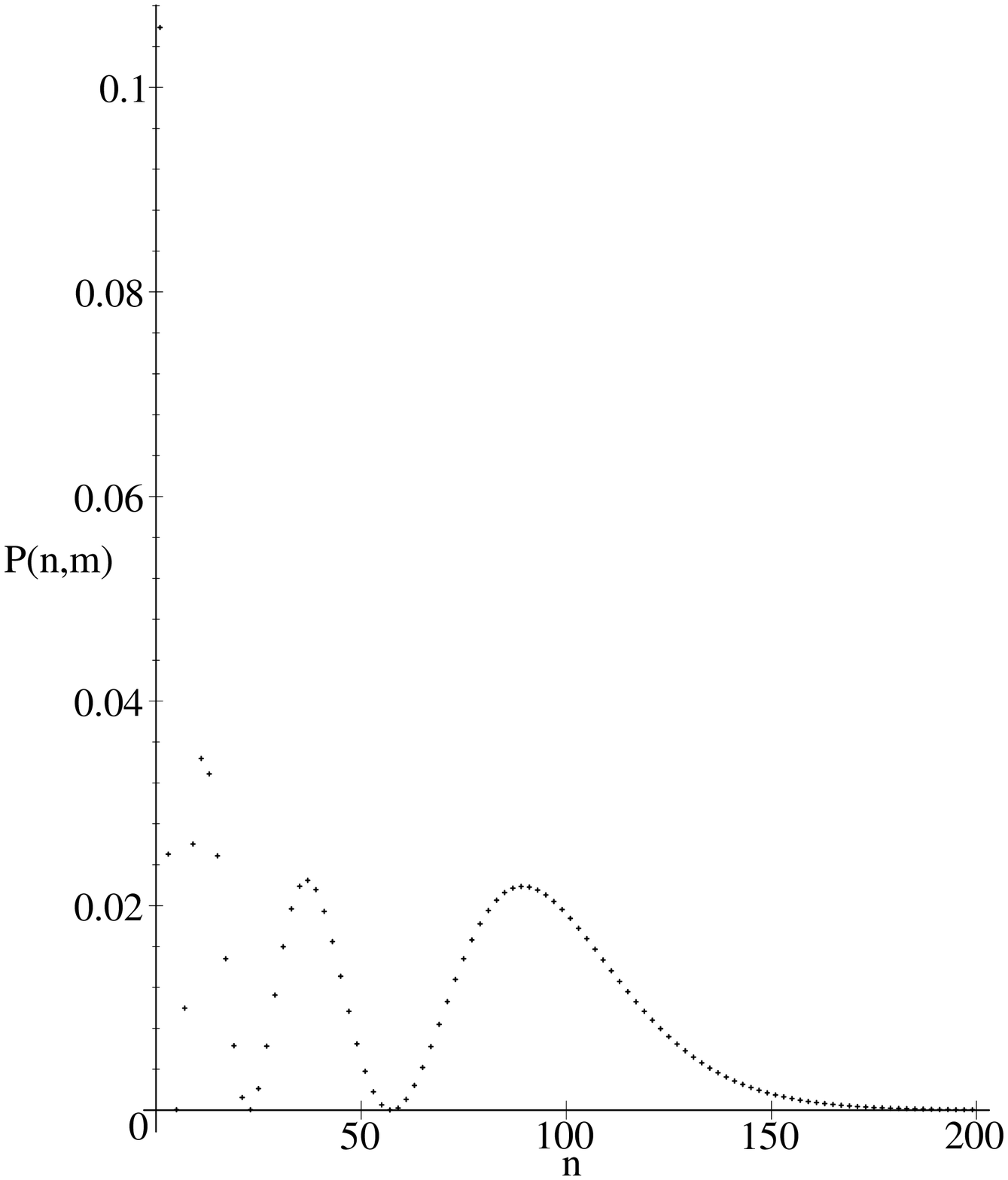}
\caption{Photon number distribution for the squeezed number states
with $m=7$ and $r=1.40$, $|\langle n|7, 1.40
\rangle|^{2}$.}\label{fig:NC7S140}
\end{figure}

 As an example, we show in Figure (\ref{fig:NC7S140})
the photon number distribution for a squeezed number state with
$m=7$ and $r=1.4$. The distribution $|\langle n|7, 1.40
\rangle|^{2}$ has 4 maxima in $n$ located at 1,11,37 and 89
photons respectively.

\begin{figure}[pt]
{\includegraphics[scale=0.5]{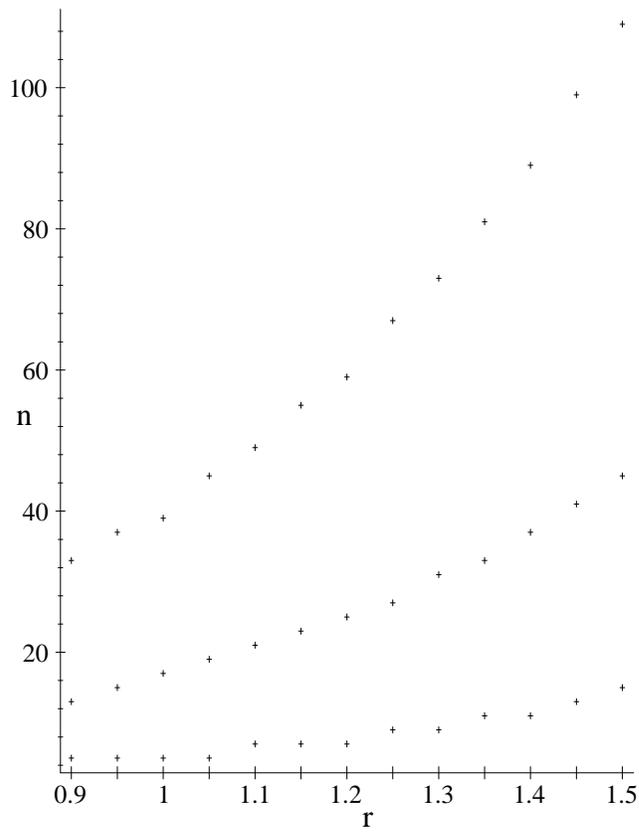}} \caption{Position
$n$ of the maxima for the photon number distribution $|\langle n|
7, r\rangle|^{2}$ of a squeezed number states, versus $r$}
\label{fig:MAX7FIG}
\end{figure}

The support of the distribution, and the oscillations, become
wider for greater $r$. For the squeezed number state with $m=7$,
this is illustrated by Figure (\ref{fig:MAX7FIG}). In this figure,
we plot the positions of $n$ for the last three maxima as
functions of $r$. These values shift to the right as $r$ grows. It
is possible to observe a tendency for each of the maxima, which
can be approximated by exponential functions .

Let us study the other representations of squeezed number states.
Using (\ref{eq:TCSPOS}) and (\ref{eq:GENESNS} ) we have
\begin{equation}\label{eq:squeezed number statesQGE}
\langle q |m,r \rangle = m!^{-1/2} (2\pi)^{-1/4}
(2^{-1/2}e^{-r})^{-1/2}
\left.\frac{\partial^{m}}{\partial\beta^{m}}
\left\{e^{\left(\frac{-(q-\sqrt{2} e^r \beta )^{2}}{2
e^{-2r} }\right)} e^{\beta^{2}/2}\right\}\right|_{\beta=0} .
\end{equation}
And using again the Cauchy formula we have
\begin{equation} \label{eq:squeezed number statesQGD}
\frac{\partial^{m}}{\partial\beta^{m}}\left\{e^{\left(\frac{-(q-\sqrt{2}e^r
\beta)^{2}}{2 e^{-2r} }\right)}e^{\beta^{2}/2}\right\}
=\sum_{k=0}^{m}\frac{m!}{(m-k)!k!}\frac{\partial^{k}}{\partial\beta^{k}}e^{\left
(\frac{-(q-\sqrt{2} e^r \beta)^{2}}{2
e^{-2r}}\right)}\frac{\partial^{m-k}}{\partial\beta^{m-k}}e^{\beta^{2}/2} .
\end{equation}
Evaluating at $\beta=0$ we obtain
\begin{eqnarray}
\langle q | m,r \rangle =
\pi^{-1/4}e^{r/2}e^{\left(-\frac{e^{2r}q^{2}}{2 }\right)} 2^{-m/2}
m!^{1/2}
\sum_{k}^{m}\frac{2^{k/2}}{k!((m-k)/2)!}\mathrm{H}_{k}\left(\frac{e^{r}q
}{2^{1/2} }\right) \label{eq:SNSFUQ}\\ \nonumber\\
k=\left\{\begin{array}{lc} 0,2,4,6... & m\mbox{ even} \nonumber\\
1,3,5,7... & m\mbox{ odd}
\end{array} \right. \nonumber
\end{eqnarray}

In order to simplify the expression above we use the following
identity of the Hermite polynomials,
\begin{eqnarray}
\frac{1}{m!}\mathrm{H}_{m}(x)=
\sum_{k}^{m}\frac{2^{k/2}}{k!((m-k)/2)!}\mathrm{H}_{k}\left(\frac{x}{2^{1/2}
}\right)\label{eq:HERMINCOG}
\\ k=\left\{\begin{array}{lc} 0,2,4,6... & m\mbox{ even} \\
1,3,5,7... & m\mbox{ odd}
\end{array} \right. \nonumber
\end{eqnarray}

Finally, the amplitude (\ref{eq:SNSFUQ}) becomes:
\begin{equation}
\langle q | m,r \rangle= \pi^{-1/4}e^{r/2}
e^{\left(-\frac{e^{2r}q^{2}}{2 }\right)} 2^{-m/2} m!^{-1/2}
\mathrm{H}_{m}\left(\frac{q}{e^{-r}}\right) \label{eq:squeezed
number statesFUQ2}
\end{equation}
Thus, the amplitude is that of a Fock state, depending on a
squeezed quadrature variable $e^{r}q$. Moreover, the amplitude of
a Fock state is given by the limit of (~\ref{eq:squeezed number
statesFUQ2}) when $r\rightarrow 0$.

The representation  of the squeezed number states in terms of the other
quadrature follows along the same lines. We get
\begin{eqnarray}
\langle p |m,r \rangle=
\frac{e^{-r/2}}{m!^{1/2}\pi^{1/4}}e^{\left(-\frac{e^{-2r}p^{2}}{2}\right)}2^{-m/2}
\sum_{k}^{m}\frac{m!(-2ie^{-r}p
)^{k}}{k!((m-k)/2)!}\label{eq:SNSFUP}\\ k=\left\{\begin{array}{lc}
0,2,4,6... & m\mbox{ even} \\ 1,3,5,7... & m\mbox{ odd}
\end{array} \right. \nonumber
\end{eqnarray}
and
\begin{equation}
\langle p |m,r \rangle=
\frac{e^{-r/2}}{m!^{1/2}\pi^{1/4}}e^{\left(-\frac{e^{-2r}p^{2}}{2}\right)}2^{-m/2}
(-i)^{m}\mathrm{H}_{m}(e^{-r}p) \label{eq:SNSFUP2}
\end{equation}

\begin{figure}[pt]
{\includegraphics[scale=0.5]{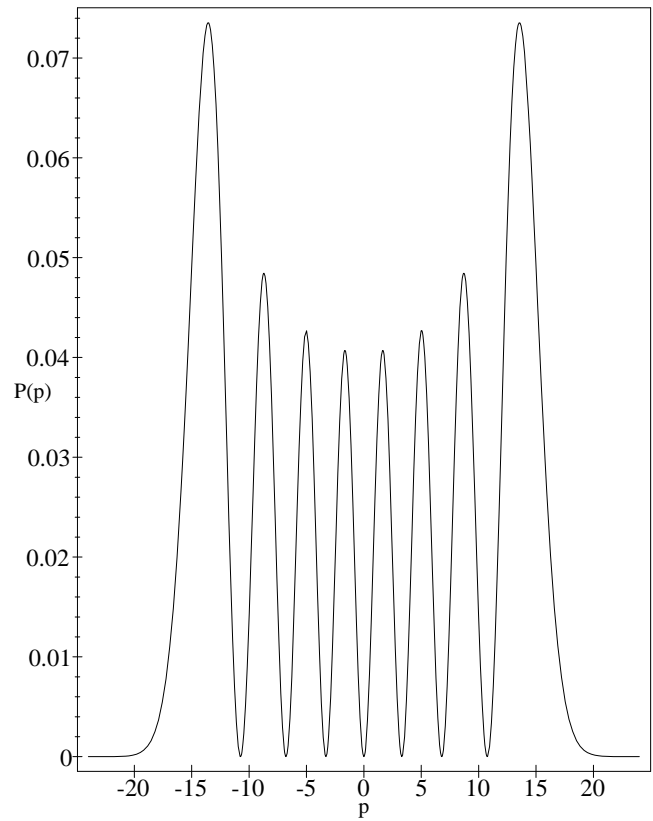}} \caption{Momentum
probability distribution $|\langle p |m,r \rangle|^{2}$ for the
$m=7$ squeezed number state with $r=1.4$.}\label{fig:P7S140}
\end{figure}

The amplitude depends on $e^{-r}p$ and, for $r>0$, $|\langle p
|m,r \rangle|^{2}$ has a support that is $e^{r}$ wider than the
support of a Fock state. The distribution has $m+1$ maxima and $m$
minima (which are also zeroes). In Figure (\ref{fig:P7S140}) we
show the momentum probability distribution for the $m=7$ squeezed
number state with $r=1.4$. The oscillations in this distribution
come from the Hermite polynomial. For large squeezing, these
oscillations reappear in the $Q(\alpha)$ distribution function.

Finally, let us turn to the $Q(\alpha)$ function, which is given
by the diagonal elements of the density matrix $\rho$ associated
with the state in the coherent states basis.

For a squeezed number state one has,
\begin{equation}
\label{eq:QFUNC}
Q(\alpha)=\frac{1}{\pi}\langle\alpha|\rho|\alpha\rangle=\frac{1}{\pi}
|\langle\alpha|m,\xi\rangle|^{2} .
\end{equation}
The function $Q(\alpha)$ gives the probability of being in a
minimum dispersion state around an average position and momentum
proportional to $Re(\alpha)$ and $Im(\alpha)$~\cite{Milburn1}.
Following the previous procedure and using (\ref{eq:TCSCOH}), we
have
\begin{equation}
\label{eq:ENCCOGE}
\langle \alpha | m , r \rangle=
\left(\frac{1}{m!\cosh(r)}\right)^{1/2}e^{-1/2|\alpha|^{2}-\tanh(r)\alpha^{*
2}/2}\left.\frac{\partial^{m}}{\partial\beta^{m}}e^{\frac{\tanh(r)}{2}\beta^{2}+\frac{\alpha^{*}}{\cosh(r)}\beta}
\right|_{\beta=0}
\end{equation}
which can be reduced to
\begin{eqnarray}
\langle \alpha|m,r \rangle=
m!^{1/2}\cosh(r)^{-1/2}e^{-1/2|\alpha|^{2}-\tanh(r)\alpha^{*2}/2}\nonumber\\
\times \sum_{p=0}^{[m/2]}
\frac{2^{-p}\sinh^{p}(r)\cosh^{p-m}(r)}{(m-2p)!p!}(\alpha^{*})^{m-2p} .
\label{eq:ENCFUCO}\\
\mbox{} [m/2]=\left \{ \begin{array}{cc} m/2 & m \quad {\rm
even} \\ \frac{m-1}{2} & m \quad {\rm odd}
\end{array}\right. \nonumber
\end{eqnarray}

The $Q(\alpha)$ function of  a squeezed number state is,
therefore:
\begin{eqnarray}
Q(\alpha)= \frac{1}{\pi}m!
\cosh(r)^{-1}e^{-|\alpha|^{2}-\tanh(r)(\alpha^{*2}+\alpha^{2})/2}\nonumber\\
\times \left| \sum_{p=0}^{[m/2]}
\frac{2^{-p}\sinh^{p}(r)\cosh^{p-m}(r)}{(m-2p)!p!}(\alpha^{*})^{m-2p}\right|^{2} .
\label{eq:ENCQF}\\ \mbox{} [m/2]=\left \{ \begin{array}{cc}
m/2 & m \quad {\rm even} \\ \frac{m-1}{2} & m \quad {\rm
odd}
\end{array}\right. \nonumber
\end{eqnarray}

\begin{figure}[pt]
 {\includegraphics[scale=0.6]{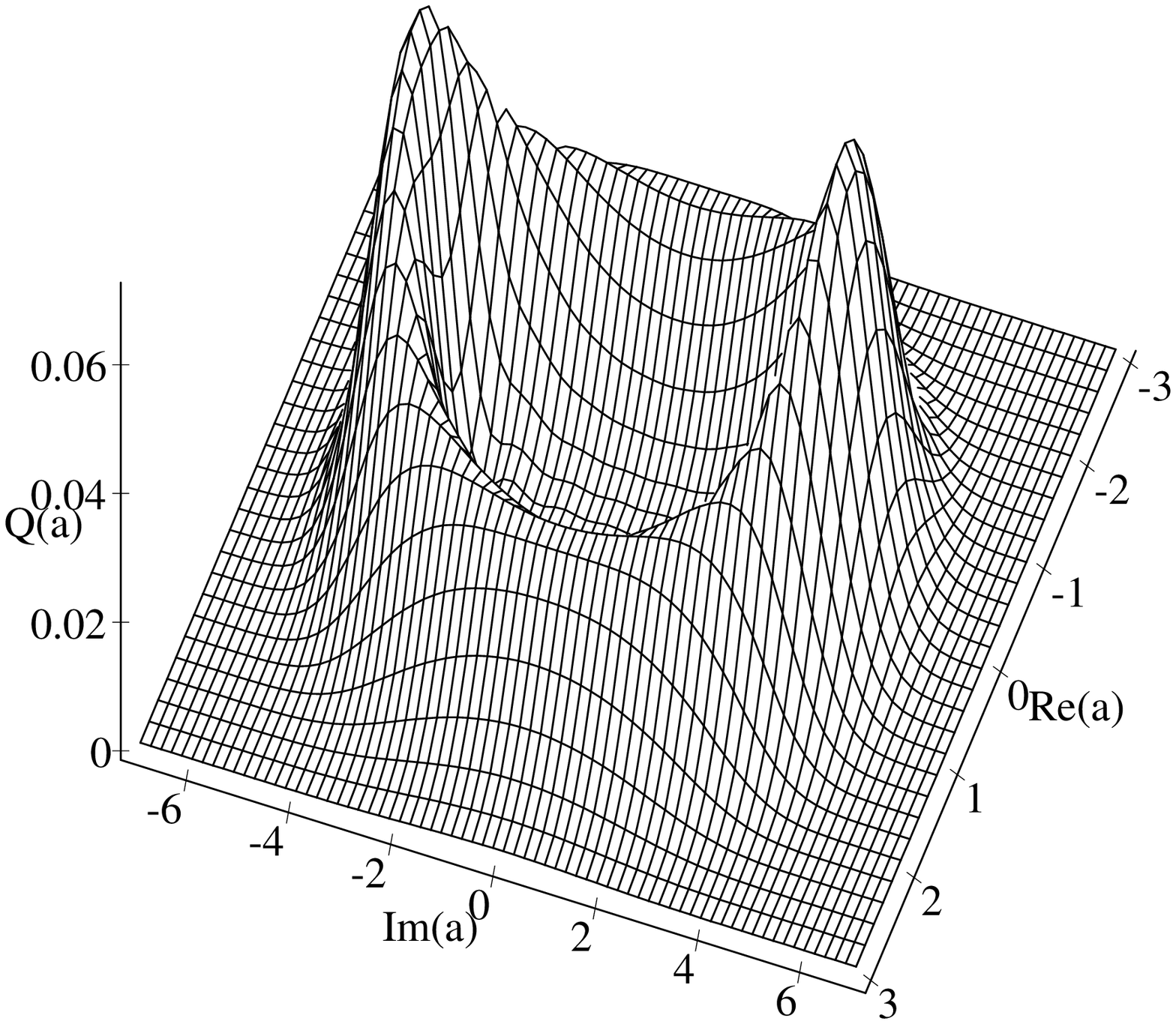}}
  \caption{$Q(\alpha)$ function of the  $m=7$  squeezed number state with
$r=0.50$}\label{fig:CN7S050}
\end{figure}

\begin{figure}[pt]
{\includegraphics[scale=0.5]{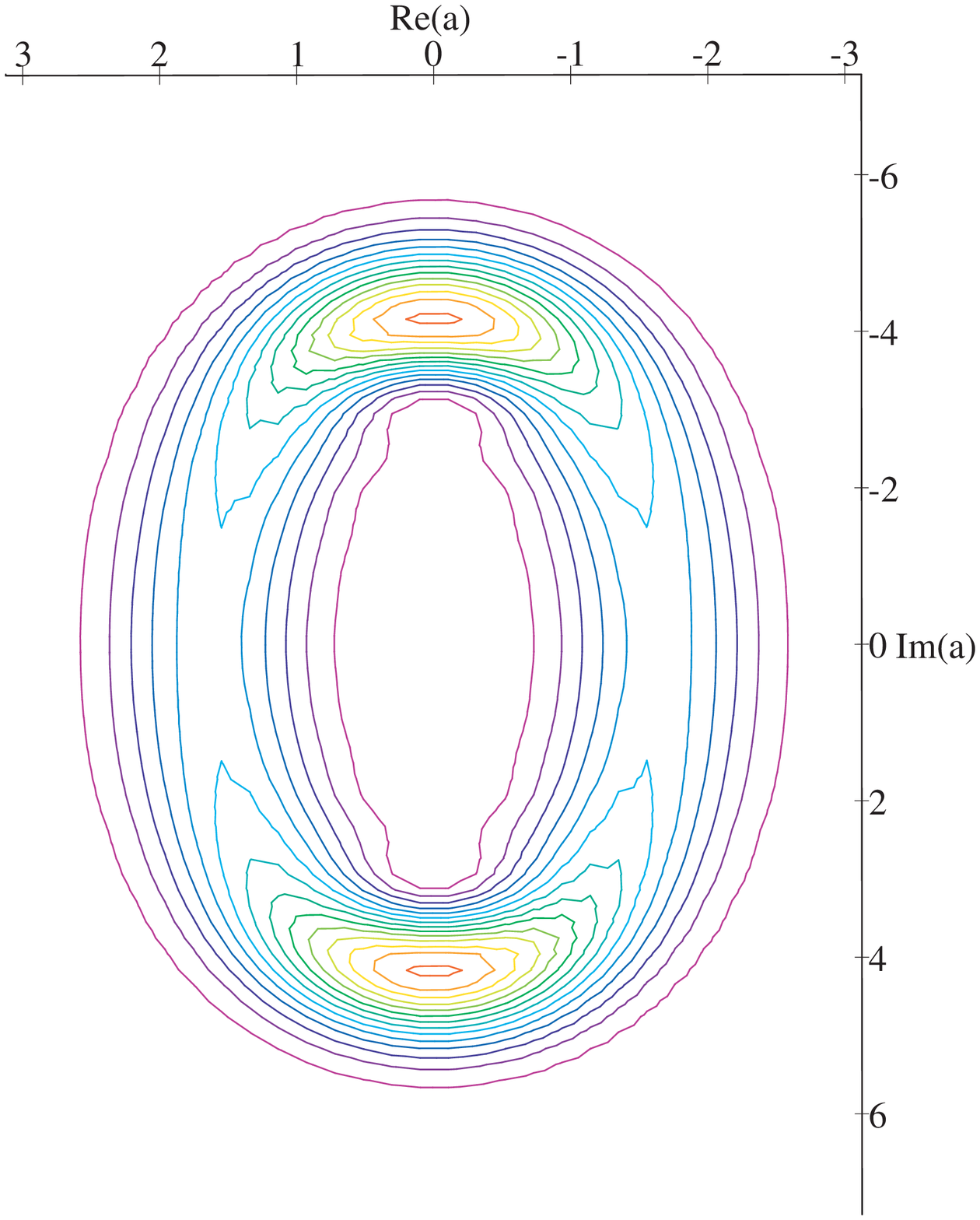}}
  \caption{Contour graphic of the $Q(\alpha)$ function of the $m=7$
  squeezed number state with $r=0.50$}\label{fig:CN7S050B}
\end{figure}

\begin{figure}[pt]
  {\includegraphics[scale=0.6]{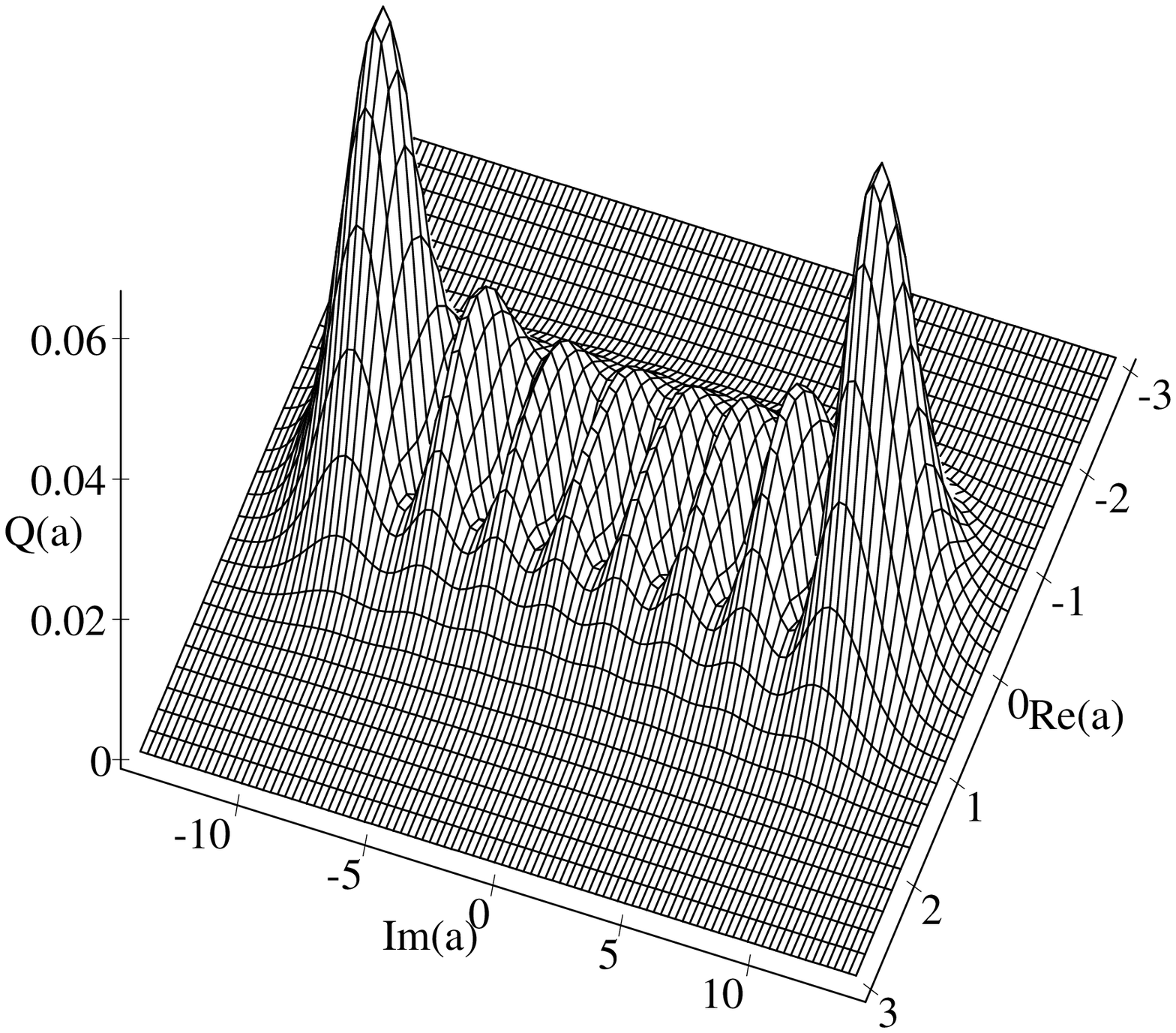}}
\caption{$Q(\alpha)$ function of the $m=7$ squeezed number state
with $r=1.40$}\label{fig:CN7S140}
\end{figure}

\begin{figure}[pt]
{\includegraphics[scale=0.5]{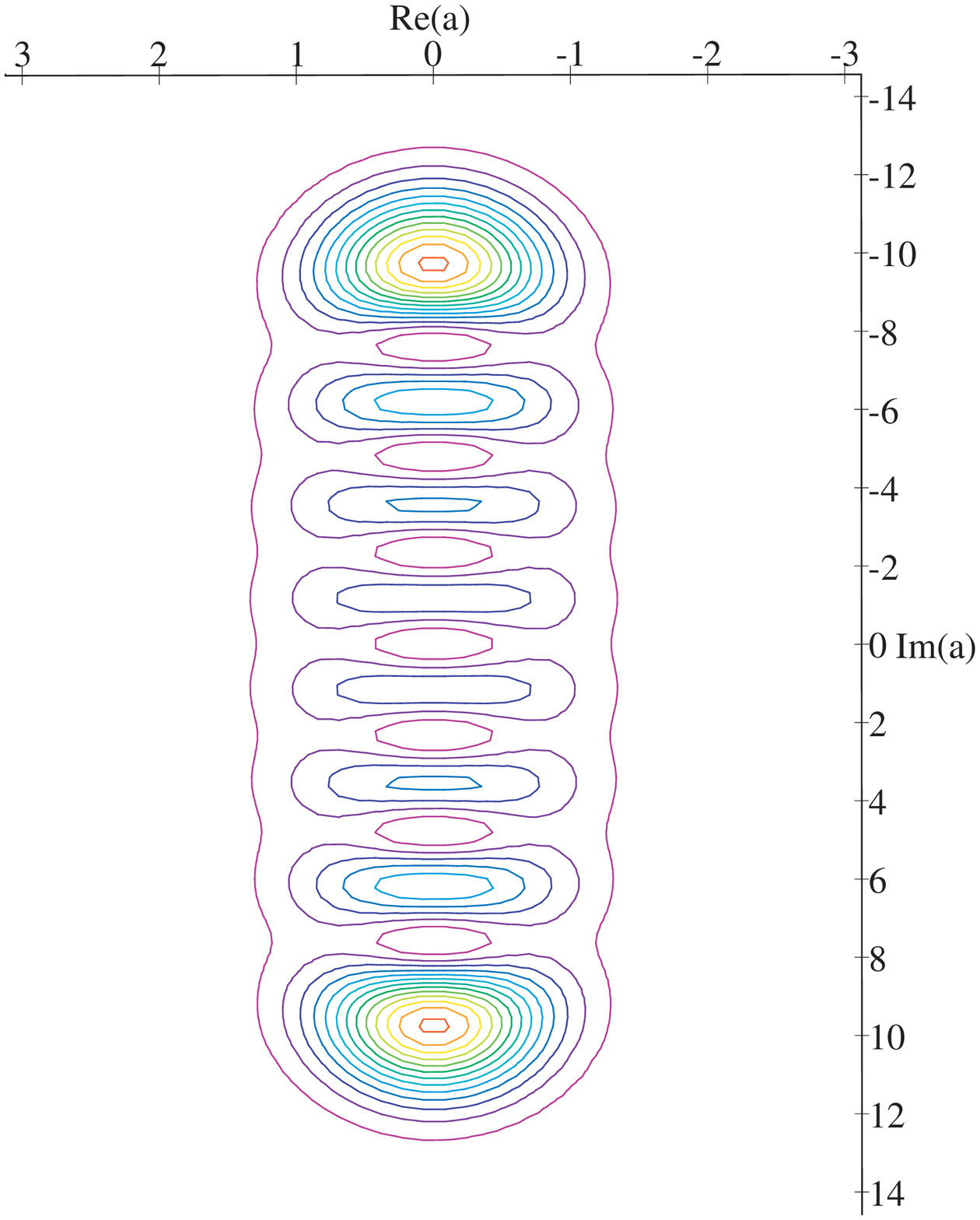}} \caption{Contour
graphic of the $Q(\alpha)$ function of the $m=7$ squeezed number
state with $r=1.40$ }\label{fig:CN7S140B}
\end{figure}

The shape of $Q(\alpha)$ depends strongly on the squeezing parameter $r$,
as illustrated by Figures (\ref{fig:CN7S050}),
(\ref{fig:CN7S050B}) and (\ref{fig:CN7S140}),
(\ref{fig:CN7S140B}); which show the Eq. (\ref{eq:ENCQF}) for
$m=7$, $r=0.50$ and $r=1.40$.

For small squeezing, the $Q(\alpha)$ function is appreciably
nonzero over an elliptical ring, with an eccentricity $e^{r}$.
When $r=0$ this ring is circular, the $Q(\alpha)$ function being
that of a Fock state. For greater squeezing , the $Q(\alpha)$
function shows prominent oscillations with a number of maxima
which for the $m$ squeezed number state stabilize at $m+1$. As can
be shown by direct numerical computation, the transition occurs at
$r\simeq \frac{1}{2}\ln(m)$ which corresponds to
$\sqrt{m}e^{-r}=1$. It is interesting to note that in the phase
space analysis  this behavior appears at the point where the inner
boundary of the deformed ring touch itself and in this sense is
also a consequence of interference in the phase space. Moreover it
can be checked by direct numerical computation that this effect is
registered only by the $Q$ function, and not by other interesting
distributions like the Wigner or Wigner-Cohen functions.

The oscillations for $r> \frac{1}{2}\ln(m)$ occur along the axis
of $Im(\alpha)$ and are correlated to the oscillations of the
momentum wave function.To show this explicitly observe that for
high squeezing, the $Q(\alpha)$ function evaluated on the
$Im(\alpha)$ axis is proportional to the square of the momentum
wave function:
\begin{equation}
Q(\alpha=i\sqrt{2}p)\sim\mid\langle p |m,r \rangle\mid^2
\end{equation}
For illustration, consider the $m=7$ squeezed number state with
$r=1.4$. The maxima in its momentum distribution (see Figure
(\ref{fig:P7S140})) correspond to those of the $Q(\alpha)$
function (figure \ref{fig:CN7S140}) in the $Im(\alpha)$ axis.

In the same way, for large squeezing, the $Q(\alpha)$ oscillations
are also correlated to the oscillations of the photon number
distribution $|\langle n| m,r \rangle|^{2}$. Each maximum $
\alpha_{max}=  i \,|\alpha_{max}|$ ($|\alpha_{max}|\geq 1$) of $Q$
is associated to a maximum $n_{max}$ of $|\langle n| m,r
\rangle|^{2}$, in such away that,
\begin{displaymath}
n_{max} \simeq |\alpha_{max}|^2 .
\end{displaymath}

In this form we can use the $Q$ function to obtain directly from
the phase space a good picture of $P_m$ distribution of the state.

\begin{figure}[pt]
{\includegraphics[scale=0.5]{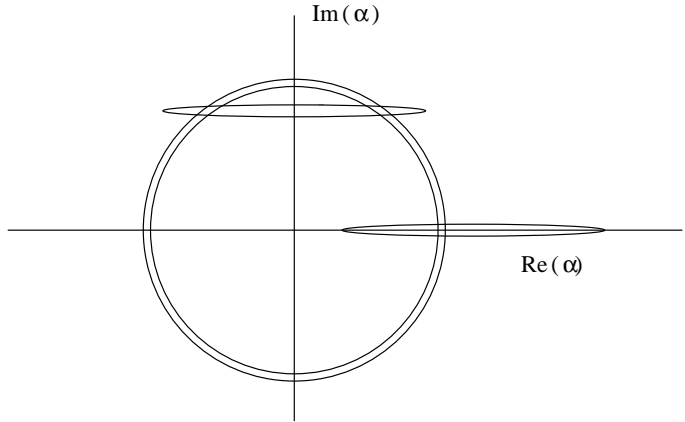}} \caption{Phase space
areas for: a number state $|m>$ (circular ring), a two photon
coherent states $S^{\dagger}|\alpha>$ with real $\alpha$ ( right
bottom ellipse ) and a two photon coherent states with imaginary
$\alpha$ (center top ellipse)} \label{dfm1}
\end{figure}

\begin{figure}[pt]
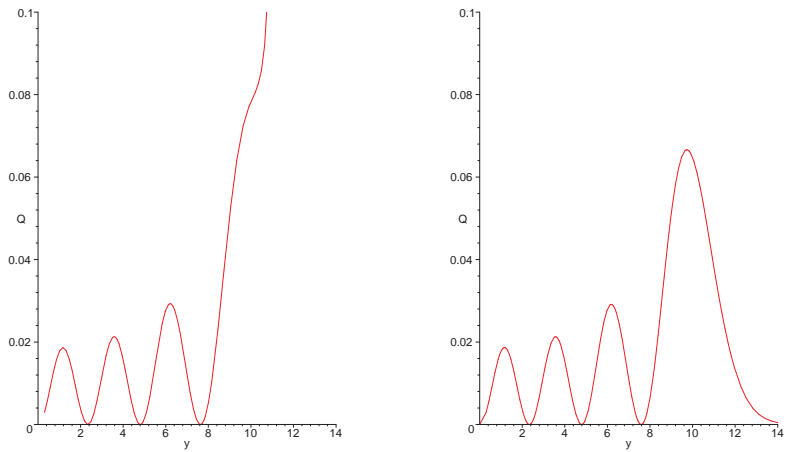

{\includegraphics[scale=0.3]{fig/osci1.ps}\includegraphics[scale=0.3]{fig/osci2.ps}}
\caption{  $Q(\alpha = i y)$ function for the $m=7$, $r=1.4$
squeezed number state: The two areas overlapping approximation (
left) and the exact calculation ( right).} \label{osci1}
\end{figure}

Finally, observe  that the $Q$ distribution of a squeezed number
state is given by ,
\begin{displaymath}
Q(\alpha) =\frac{1}{\pi}|<\alpha| S |m> |^2=\frac{1}{\pi}|<m| S^{\dagger}
|\alpha> |^2 .
\end{displaymath}
and henceforth, the $Q(\alpha)$ function of a squeezed number
state $| m,r \rangle$ is proportional to the photon number
amplitude of a two photon coherent state $| \alpha,r \rangle=
S^{\dagger} |\alpha>$. The oscillations still can  be explained
using the hypothesis of interference in phase space. To illustrate
this point consider Figure (\ref{dfm1}) which shows the areas
representing a Fock state and a pair of two photon coherent states
( $| \alpha,r \rangle$) with respectively real or imaginary
$\alpha$ . For $\alpha $ imaginary the overlapping area doubles
and produces the interference effect in the $Q$ function. The
photon statistical value of a two photon coherent state is
computed using the method of overlapping areas in the phase space
\cite{Schleich3}. The result is,
\begin{equation}
P_m = |<m|S^{\dagger}|\alpha = i y>|^2 = |\sqrt{Am} e^{i \phi} + \sqrt{Am} e^{-i \phi} |^2
\end{equation}
with
\begin{equation}
A_m = \frac{e^{-2(m+1/2 -y^2/e^{2r}) / e^{2r}}}{\sqrt{ 2 \pi ( m+1/2 -y^2/e^{2r} )  }}
\end{equation}
and
\begin{eqnarray}
\phi& =& (m+1/2) \arctan \{ \sqrt{( m+1/2 -y^2/e^{2r}/(y^2/e^{2r}})\}\nonumber\\
&&-x \sqrt{( m+1/2 -y^2/e^{2r}/e^{2r}})-\pi/4
\end{eqnarray}
This match  with the exact computation presented above as is shown
in Figure (\ref{osci1}) where we present respectively the
$Q(\alpha= iy)$ function for the $m=7$, $r=1.4$ squeezed number
state obtained using this approximation and the previous exact
calculation. For sufficiently high values of $|\alpha|$ the WKB
approximation on which is based the phase space computation no
longer holds as evidenced in Figure (\ref{osci1}).

\section{Conclusion}

We have computed the photon number distribution, the momentum like
and position like wave functions, and the $Q(\alpha)$ function for
squeezed number states and we have shown that each of them has
characteristic oscillations which depend on the squeeze parameter
$r$. For highly squeezed number states we observed that the
oscillations in the different probability distributions are fixed
in number and are in close correspondence between them. The
$Q(\alpha)$ function in this case present a richer structure that
for  number states. The oscillations were also discussed within
the semiclassical approximation using the phase space methods
based in the WKB functions \cite{Dowling, Schleich3}.

It should be also mentioned that the generating function formalism
used throughout this work has been shown to be a straightforward
and very valuable procedure for calculating probability amplitudes
and matrix elements for any base obtained transforming number
states by any arbitrarily chosen unitary operator.

\section{ Acknowledgements}

We thank a referee for important observations on an earlier
version of this article. This work was supported by DID-USB and by
Grant N0 G-2001000712 of Fonacit.

\newpage

\end{document}